\documentclass[11pt]{article}

\usepackage[letterpaper,margin=1in]{geometry}
\usepackage{microtype}
\usepackage{amsmath,amssymb}
\usepackage{booktabs}
\usepackage{siunitx}
\usepackage{graphicx}
\usepackage{subcaption}
\usepackage{longtable}
\usepackage{array}
\usepackage{setspace}
\usepackage{enumitem}
\usepackage[colorlinks=true,allcolors=blue]{hyperref}
\usepackage{float}
\usepackage[english]{babel}
\usepackage{chngcntr}
\usepackage{tikz}
\usetikzlibrary{positioning, arrows.meta, fit, calc, shapes.geometric}

\graphicspath{{./}{./figures/}}
\setlist[itemize]{leftmargin=*,itemsep=0.25em,topsep=0.25em}
\setlist[enumerate]{leftmargin=*,itemsep=0.25em,topsep=0.25em}
\newcommand{\secref}[1]{\hyperref[#1]{Section~\ref*{#1}}}
\newcommand{\subsecref}[1]{\hyperref[#1]{Section~\ref*{#1}}}
\newcommand{\figref}[1]{\hyperref[#1]{Figure~\ref*{#1}}}
\newcommand{\tabref}[1]{\hyperref[#1]{Table~\ref*{#1}}}

\title{High-Volume Plaintiff-Side Counsel and Single-Appearance Eviction Cases in Philadelphia}
\author{
  Marios Papamichalis\thanks{Human Nature Lab, Yale University, New Haven, CT 06511, \texttt{marios.papamichalis@yale.edu}}
  \and
  Regina Ruane\thanks{Department of Statistics and Data Science, The Wharton School, University of Pennsylvania, 3733 Spruce Street, Philadelphia, PA 19104-6340, \texttt{ruanej@wharton.upenn.edu}}
}
\date{}

\begin{document}
\maketitle

\begin{abstract}
Among 755{,}004 Philadelphia landlord--tenant records filed during 1969--2022, 396{,}163 residential cases involve tenants who appear exactly once in the observed docket. In unadjusted comparisons, single-appearance cases handled by high-volume plaintiff-side counsel are more likely to advance to the writ-of-possession and served-writ stages, but no more likely to end in default. Comparisons within the same plaintiff, and within the same plaintiff at the same property, show no broad premium on adverse case outcomes such as default, judgment, or fees. The clearer pattern is organizational: after a plaintiff adopts or switches into high-volume counsel, monthly filings rise by about 2--5\% and the number of distinct buildings reached rises by a similar margin; near the prior-year top-10 attorney threshold, cases display local differences in default and enforcement; and continuances under specialist counsel are more closely linked to default. Non-flat pre-treatment trends and imprecise reverse-direction estimates from attorney exits restrict the strength of any causal claim. High-volume plaintiff-side counsel therefore functions as a mechanism of filing scale and procedural sequence, not as a uniform escalator of case outcomes or as a cause of any individual tenant becoming single-appearance.
\end{abstract}

\noindent\textbf{Keywords:} eviction court; legal intermediation; staggered difference-in-differences; Callaway--Sant'Anna; high-dimensional fixed effects; administrative court records.

\noindent\textbf{arXiv subject classification:} Statistics - Applications (stat.AP).

\section*{Significance Statement}

Eviction research typically counts filings or judgments. The relevant population here is narrower: residential cases in which the tenant key appears exactly once in the full observed Philadelphia landlord--tenant docket. Across this population, specialist cases are more likely to reach writ and served-writ stages in unadjusted comparisons, but within-unit comparisons show no stable premium on adverse endpoints. The clearer signal is scale and sequence: filing volume and building reach rise after adoption of, or switching into, high-volume counsel; local threshold comparisons reveal selective enforcement differences; and continuances under specialist counsel are more default-linked. The causal estimand is plaintiff-month initiation of filings into the single-appearance population, not whether counsel causes a particular tenant to become single-appearance.

\section{Introduction}\label{sec:introduction}

The Philadelphia landlord--tenant docket contains a large population of cases in which the tenant key appears exactly once: residential filings against tenants who never reappear in the observed 1969--2022 record. These are referred to here as \emph{single-appearance tenant cases}. The definition is administrative and does not capture informal housing conflict, displacement outside court, or executed eviction by writ; it identifies formal court entries for tenants observed exactly once in the docket.

The empirical question is how high-volume plaintiff-side counsel organizes these cases. One account predicts a broad adverse-outcome premium: specialist counsel should raise default, judgment, writ, served-writ, fee, and award measures across the board. A second account predicts scale and sequence: specialist counsel may expand where and how often plaintiffs file, and may change the procedural route through the docket, without uniformly worsening every endpoint after filing.

Unadjusted comparisons cannot distinguish these accounts because specialist cases differ in plaintiff, property, tenant, judge, and filing-period composition. To separate stable statistical patterns from compositional differences, the analysis combines unadjusted outcome comparisons with within-plaintiff and within-plaintiff--property models, staggered adoption and switch designs, local rank-threshold comparisons, placebo checks, and follow-up-balanced robustness.

The dominant pattern is scale and sequence. Unadjusted specialist-counsel cases are substantially more likely to advance to writ-of-possession and served-writ stages, but no more likely to end in default. Within-unit comparisons show no broad premium on adverse outcomes, and the share of award attributable to fees and costs does not rise once the same plaintiff is followed over time. Adoption and switch designs yield significant post-treatment increases in plaintiff-month filing volume and the number of distinct buildings reached. Local rank-threshold estimates near the prior-year top-10 attorney cutoff show significant differences in default and enforcement margins, and continuances under specialist counsel are more tightly linked to default.

The scope of interpretation is narrow. Adoption and switching into high-volume counsel are not randomized, and pre-treatment trend diagnostics indicate that plaintiffs often begin to change behavior before treatment is recorded. The adoption and switch results therefore identify post-treatment scale changes under conditional parallel trends rather than randomized treatment effects. Single-appearance status is itself a docket-wide classification, so the causal estimand is plaintiff-month initiation of filings into that population, not an individual tenant outcome.

The findings characterize plaintiff-side legal intermediation in a large single-appearance eviction population. High-volume counsel functions primarily as a mechanism for filing throughput, geographic reach, and procedural routing, rather than as a uniform escalator of case-level outcomes.

Section~\ref{sec:literature} situates the analysis in the eviction and civil-court literatures. Section~\ref{sec:data} describes the data, measurement windows, and empirical strategy. Section~\ref{sec:results} reports descriptive, within-unit, staggered-adoption, threshold, and robustness results. Section~\ref{sec:discussion} concludes.

\section{Literature}\label{sec:literature}

The eviction literature is large but mostly addresses prevalence and harm. The narrower question pursued here is whether specialist plaintiff-side counsel changes how cases are organized when tenants appear exactly once in the formal eviction docket.

Recent eviction scholarship establishes why that single-appearance question matters. Eviction is not an idiosyncratic response to isolated tenant default, but a regular legal mechanism through which poverty, hardship, instability, and downstream harms are reproduced \cite{hartman2003evictions,desmond2012eviction,desmond2017evicted}. Related work documents health, residential-instability, neighborhood, and discrimination-related consequences \cite{desmond2015eviction,desmond2017gets,desmond2015forced,greenberg2016discrimination}. Large administrative-record projects and linked causal studies have clarified both the scale of formal eviction and the stakes of court-based housing instability \cite{gromis2022estimating,graetz2023comprehensive,collinson2024eviction}. That literature also shows why eviction cannot be treated as a single binary event. Filings, judgments, writs, and executed removals are not interchangeable legal stages; post-judgment enforcement is often incompletely observed; court records contain boundary and recording problems; and formal dockets miss a substantial share of informal displacement \cite{nelson2021evictions,porton2021inaccuracies,zainulbhai2022informal}. Related work on eviction-court displacement rates further shows that the share of filed cases ending in actual court-ordered displacement varies sharply across jurisdictions, reinforcing the need to distinguish first court entry from later enforcement \cite{summers2022eviction}. Those measurement lessons are directly relevant here because the analysis studies single-appearance court contact as a legal ladder---default, judgment, writ, served writ, fee burden, and judgment by agreement---rather than as a filing alone, and because the Philadelphia dockets link those stages inside the same record \cite{gromis2022estimating,summers2025pathways,summers2022eviction}.

A second literature explains why single-appearance cases may already be institutionally filtered rather than isolated tenant disputes. Court cases are socially selected disputes, and settlement must be understood as a structured legal outcome rather than as residual noise \cite{felstiner2017emergence,mnookin1978bargaining,priest1984selection}. Repeat-player theory predicts that frequent litigants gain strategic advantage from specialization, familiarity, and repeated exposure to court routines \cite{galanter1974haves}. Work on lawyerless and high-volume courts shows that defaults and negotiated outcomes are produced in organizational settings marked by representation gaps, compressed bargaining, and limited adjudicative attention \cite{bezdek1991silence,engler2010connecting,shanahan2022judges,sabbeth2022eviction,sudeall2021praxis}. Recent ethnographic work sharpens that institutional claim by showing that tenant outcomes are shaped not only by the absence of counsel but also by informal procedural practices, scheduling pressure, efficiency norms, and courtroom credibility assessments that can affect parties asymmetrically in high-volume dockets \cite{fleming2023navigating,kepes2024system,hanley2024power}. Studies of tenant counsel likewise show that attorneys can affect outcomes, even if the size and mechanism of those effects vary across settings \cite{seron2001impact,ellen2021lawyers,cassidy2023effects,summers2024evicted}. Newer work on tenant right-to-counsel implementation further suggests that defense-side capacity should not be treated as a simple present/absent variable: RTC laws vary in trigger point, eligibility, and administration, and tenants can still face substantial learning, compliance, and psychological burdens in accessing counsel even where programs formally exist \cite{benfer2025descriptive,von2025no}. More recent work on assembly-line plaintiffs and professional evicting attorneys extends that insight to the plaintiff side, showing that specialist landlord counsel can itself become a scalable market layer \cite{aizman2025shadow}. Evidence that procedural frictions such as travel burden raise default, while more time to disposition and tenant representation improve tenants' odds of avoiding an eviction judgment, further supports treating timing and sequence as institutional mechanisms rather than administrative residue \cite{ryan2024buying}. More generally, courtroom-workgroup research suggests that repeated actor familiarity can shape both disposition mode and time to disposition, providing a broader organizational frame for the attorney specialization examined here \cite{metcalfe2016role}. At the same time, quasi-experimental eviction studies underscore how difficult strong causal identification can be in housing-court settings, motivating the use of within-plaintiff, within-property, and adoption-linked comparisons rather than unadjusted specialist versus non-specialist contrasts alone \cite{collinson2024eviction}.

A third literature on landlord--tenant law, legal expertise, and legal infrastructure helps explain the present focus on sequencing, fee burden, and courtroom routing rather than on judgments alone. Modern landlord--tenant governance increasingly turns on standardized, contract-centered legal forms rather than older status-based arrangements \cite{rabin1983revolution}. Legal expertise in court is also relational and procedural, not merely doctrinal \cite{sandefur2015elements}. Law-in-action scholarship likewise cautions against treating formal rights as self-executing: even tenants with meritorious claims often fail to benefit from protective doctrine in practice, underscoring how strongly substantive law is filtered through institutional process \cite{summers1limits}. More broadly, legal infrastructure includes not only formal rules, but also the organized intermediaries and recurring pathways through which law structures bargaining power and allocates risk \cite{pistor2019code,sandefur2015elements}. That framing matters here because, in single-appearance eviction, specialist counsel functions as a technology of throughput and procedural sequence inside an uneven legal field, rather than as a uniform escalator of case-level outcomes.

A fourth literature shows why those mechanisms matter substantively. Research on serial filing and large-owner strategy demonstrates that landlords often use court as a repeat rent-collection and tenant-discipline technology rather than only as a direct route to physical removal \cite{garboden2019serial,leung2021serial,immergluck2020evictions}. Complementing that literature, newer landlord-side evidence suggests that larger landlords use eviction more as a standardized loss-recovery technology and are substantially more likely than smaller landlords to proceed to final judgment, consistent with an organizational reading of the scaling results reported below \cite{ryan2025tale}. National demographic work and disparity studies show that exposure to eviction is highly unequal by race and gender \cite{graetz2023comprehensive,hepburn2020racial}. Related work on filing fees, civil extraction, and legal debt shows that court-generated charges can reshape filing incentives, redistribute monetary burdens to tenants, and deepen inequality \cite{gomory2023racially,ajayi2026landlord,brito2022racial,harris2010drawing}. These studies motivate sustained attention to single-appearance scale, debt, and the possibility that specialist counsel organizes single-appearance eviction through process and bargaining even where it does not uniformly raise downstream endpoints.

A fifth literature shows that settlement outcomes conceal a substantive contractual layer with consequences that outlast the hearing itself. Eviction agreements can operate through probation-like obligations, standardized adhesion terms, and heterogeneous enforcement pathways that are not visible in coarse disposition codes \cite{summers2023civil,summers2026settlements,summers2025pathways}. Newer work on courthouse bargaining likewise suggests that courts do not merely record settlement outcomes. Organizational rules and workgroup norms can shape the path toward bargaining in ways that make settlement itself part of the institutional process associated with later housing instability \cite{hardaway2026courthouse,summers2023civil,summers2026settlements}. Parallel work on tenant screening shows that court records can generate durable collateral consequences even without physical removal because landlords and screening firms reuse filing information in discretionary and often opaque ways \cite{kleysteuber2006tenant,eisenberg2024record,brantley2025record}. Pandemic-era work further demonstrates that eviction systems can be interrupted at particular stages, so filing, bargaining, and downstream enforcement should not be assumed to move together \cite{hepburn2023protecting,benfer2023covid}. Those insights are directly relevant here because the results below distinguish unadjusted enforcement differences from the narrower within-unit mechanisms of scale, sequence, and fee composition. A complementary study of the same Philadelphia landlord--tenant docket shows how plaintiff-side concentration, repeat-address filing, tenant recurrence, and specialist plaintiff counsel organize the broader residential eviction system \cite{papamichalis2026legal}.

Taken together, these literatures leave a specific gap. Much existing work explains who is evicted, how much eviction occurs, or what happens after court contact; far fewer studies isolate what specialized plaintiff-side counsel does in the large population of tenants who appear only once in the observed docket. The evidence below addresses that institutional question. Specialist plaintiff-side counsel does not uniformly intensify single-appearance tenant cases; it widens court reach, reorganizes case sequencing, and shifts fee and enforcement pathways inside an unevenly defended courtroom field.

\section{Data, measurement windows, and empirical strategy}\label{sec:data}

\subsection{Data}

The data are the Philadelphia Municipal Court landlord--tenant docket for filing years 1969--2022, with an archival universe of 755{,}004 records. The single-appearance audit identifies 517{,}272 unique tenant keys with usable tenant information; 401{,}527 appear exactly once and 115{,}745 appear more than once. The main analytic sample is the residential single-appearance subset: 396{,}163 residential cases in which the tenant key appears once in the full observed docket and never reappears.

This definition is intentionally stricter than an as-of first-observed definition. A first-observed definition counts the first filing for every tenant, including tenants who return later. The single-appearance definition used here counts only tenants observed once in the entire docket. Thus, the unit of analysis is not all first appearances by tenants. It is the subset of tenant cases that appear to be single-appearance entries in the observed Philadelphia court record.

Specialist plaintiff-side counsel appears in 212{,}074 of the 396{,}163 residential single-appearance cases, or 53.5\%. Other, non-specialist, or missing plaintiff-side counsel appears in 184{,}089 cases, or 46.5\%. These counts define the denominators for the results below.

\begin{table}[t]
\centering
\caption{Docket and single-appearance sample construction.}
\label{tab:oneshot_manifest}
\small
\begin{tabular}{p{0.48\linewidth}p{0.18\linewidth}p{0.26\linewidth}}
\toprule
Quantity & Count & Interpretation \\
\midrule
Unique tenant keys with usable tenant information & 517{,}272 & observed tenant identities in the filtered docket \\
Tenant keys seen exactly once & 401{,}527 & full-docket single-appearance tenants \\
Tenant keys seen more than once & 115{,}745 & repeat tenants in the observed docket \\
residential single-appearance rows & 396{,}163 & main analytic sample \\
Specialist / high-volume plaintiff-side attorney cases & 212{,}074 & 53.5\% of the analytic sample \\
Other, non-specialist, or missing plaintiff-side attorney cases & 184{,}089 & 46.5\% of the analytic sample \\
Rows missing usable tenant key & 104 & excluded from tenant-lifetime construction \\
\bottomrule
\end{tabular}
\end{table}

\subsection{Measurement windows}

The single-appearance definition uses the full observed docket because a tenant key can be classified as single-appearance only after it is checked across the full panel. Other modules use narrower windows when the relevant fields become informative. Plaintiff-side attorney definitions begin in 1983, when attorney identifiers become substantively stable enough to define high-volume plaintiff-side counsel. Fee and award models are modern-period models because fee fields become informative later.

\begin{table}[t]
\centering
\caption{Measurement windows by empirical module.}
\label{tab:measurement_windows}
\small
\begin{tabular}{p{0.36\linewidth}p{0.18\linewidth}p{0.36\linewidth}}
\toprule
Module & Primary window & Rationale \\
\midrule
single-appearance definition & 1969--2022 & requires checking whether each tenant key appears once in the full observed docket \\
Unadjusted single-appearance descriptive ladder & 1969--2022 & summarizes the full residential single-appearance sample \\
Specialist adoption and within-plaintiff models & 1983--2022 & requires stable plaintiff-side attorney names and lagged high-volume definitions \\
Within-plaintiff--property models & 2003--2022 in the identified cells & requires plaintiff--property cells with attorney-regime variation \\
Portfolio scaling models & 1983--2022 & uses plaintiff-month adoption timing and single-appearance filing counts \\
Fee and award models & modern fee window & fee components are substantively informative only in later years \\
\bottomrule
\end{tabular}
\end{table}

\subsection{Core measures}

The central sample indicator is single-appearance tenant status. A case is coded as single-appearance if the tenant key appears exactly once in the full observed docket. This is a data-based administrative definition. It does not mean the tenant had no informal housing conflict, and it does not mean the tenant was actually removed. It means that the tenant key appears once, and only once, in the observed Philadelphia landlord--tenant court record. Because this is an administrative tenant key rather than a verified person-level identifier, the count should be read as a count of usable tenant identities in the court record. Name variation, shared names, or matching error may split or combine real people.

The central treatment is specialist plaintiff-side counsel. Specialist counsel refers to high-volume plaintiff-side eviction attorneys. In the main specification, an attorney is coded as specialist in year $y$ if the attorney falls in the prior-year top decile of plaintiff-side filing volume. This is a data-based category, not a formal legal certification. A plaintiff is coded as having adopted specialist counsel in the first month in which the plaintiff uses such counsel.

The central outcomes are default, judgment by agreement (JBA), judgment for plaintiff, writ of possession, served writ, fee share, and award-over-sought. Served writ refers to service of the alias writ and is used as the principal observed enforcement indicator. Fee share is defined for cases with a positive total award as
\[
\mathrm{FeeShare}_i =
\frac{\mathrm{AttorneyFees}_i + \mathrm{Costs}_i + \mathrm{OtherFees}_i}{\mathrm{TotalAward}_i}.
\]
Award-over-sought is defined for cases with positive total award and positive amount sought as
\[
\mathrm{AwardOverSought}_i =
\frac{\mathrm{TotalAward}_i}{\mathrm{AmountSought}_i}.
\]

\subsection{Empirical strategy}

The empirical strategy has six components. First, unadjusted outcome ladders compare specialist and non-specialist cases in the residential single-appearance sample. Second, within-plaintiff and within-plaintiff--property models test whether unadjusted differences survive comparisons within the same filing actor or the same filing actor at the same property. Third, staggered adoption and switch designs estimate post-treatment effects on plaintiff-month filing counts and building counts. Fourth, a standard \texttt{ATTgt} package implementation at quarter resolution validates the direction of the staggered estimates. Fifth, local threshold comparisons at the prior-year rank-10 attorney cutoff isolate selection at the boundary of the high-volume category. Sixth, sequence, fee, placebo, follow-up-balanced, judge, and plaintiff--property exposure diagnostics probe selective mechanisms.

For filed cases, the main within-plaintiff design is
\[
Y_{ipt}=\alpha_p+\lambda_t+\beta\,\mathrm{Specialist}_{ipt}+X_{ipt}\gamma+\varepsilon_{ipt},
\]
and the strongest within-unit comparison is
\[
Y_{ipt}=\alpha_{p\times property}+\lambda_t+\beta\,\mathrm{Specialist}_{ipt}+X_{ipt}\gamma+\varepsilon_{ipt}.
\]
Binary outcomes are estimated with linear probability models and continuous outcomes with OLS. Standard errors are clustered at the plaintiff level where the design allows it.

The scale analysis uses a CSDID design because attorney adoption occurs at different times across plaintiffs and treatment effects may vary across cohorts. Staggered two-way fixed-effects estimators can place non-transparent or negative weights on cohort comparisons when treatment effects are heterogeneous \cite{goodman2021difference,de2020two,sun2021estimating}. We therefore estimate cohort-time average treatment effects using the Callaway--Sant'Anna comparison of each treated cohort with plaintiffs that are not yet treated or never treated \cite{callaway2021difference}. Let $G_p$ denote plaintiff $p$'s first adoption or switch month, with $G_p=\infty$ for never-treated plaintiffs. Let $Y_{pt}=\log(1+\mathrm{filings}_{pt})$ or $\log(1+\mathrm{buildings}_{pt})$. For cohort $g$ and event time $k=t-g$, the comparison is
\[
\widehat{ATT}(g,k)=
E\!\left[Y_{p,g+k}-Y_{p,g-1}\mid G_p=g\right]
-
E\!\left[Y_{p,g+k}-Y_{p,g-1}\mid p\in C_{g+k}\right],
\]
where $C_{g+k}=\{p:G_p>g+k\ \mathrm{or}\ G_p=\infty\}$ is the not-yet-treated and never-treated control set. Event-time effects are weighted by the number of treated plaintiffs contributing to each cohort-time cell. The main post-treatment summary averages event times $k=1,\ldots,12$ months. Uncertainty in the monthly tables is computed with a cell-level Rademacher multiplier bootstrap. Lead coefficients are reported as pretrend diagnostics.

Two treatment definitions are used. The top-decile adoption design defines treatment as the first month in which a plaintiff uses a prior-year top-decile plaintiff-side attorney. The top-10 switch design defines treatment as the first month in which a plaintiff uses a prior-year top-10 plaintiff-side attorney after previously filing outside that category. To make the switch design a genuine switching comparison rather than a first-observed filing comparison, CSDID estimation is restricted to plaintiffs with at least six pre-switch months and at least three pre-switch filings. This retains 1{,}047 of 73{,}445 nominal top-decile adopters and 1{,}023 of 61{,}767 nominal top-10 switchers.

As package-level validation, \texttt{differences.ATTgt} is run at quarter resolution for the same four scale specifications. The monthly CSDID output remains the main high-resolution estimator; the quarterly package run verifies sign and magnitude over the comparable post-1--4-quarter window. The package's simple aggregation is not cited, since it averages over a broader event grid and is not comparable to the near-post window. Package post-window $p$-values are treated as heuristic because adjacent event-time estimates share treated and control units.

Statistical significance is reported for unadjusted post-treatment CSDID effects, fixed-effect estimates, rank-threshold estimates, sequence estimates, and robustness estimates. Pre-treatment trend tests qualify causal inference: when pre-treatment trends reject, pretrend-adjusted estimates are reported as sensitivity quantities rather than as significant positive findings.

\section{Results}\label{sec:results}

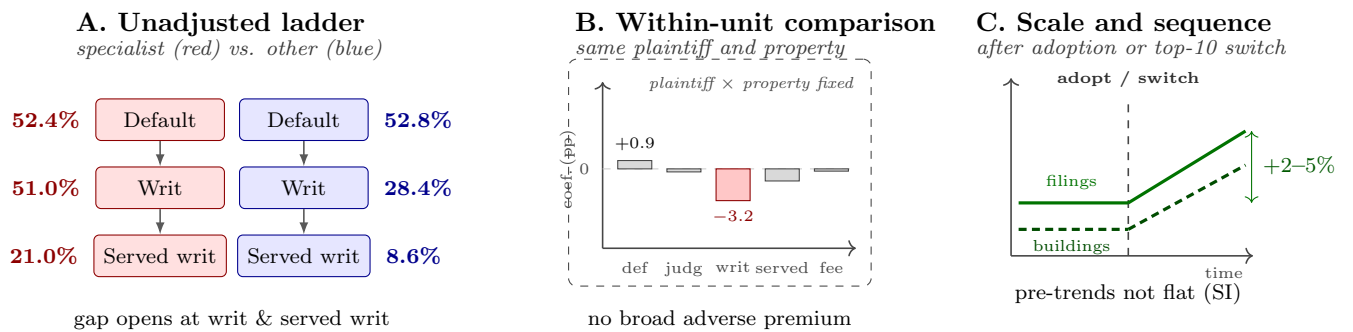
\begin{figure}[H]
\centering
\begin{tikzpicture}[
    font=\small,
    stage/.style={draw, rounded corners=2pt, minimum width=1.7cm, minimum height=0.55cm, align=center, font=\scriptsize},
    spec/.style={stage, fill=red!12, draw=red!55!black},
    nonspec/.style={stage, fill=blue!10, draw=blue!55!black},
    arr/.style={-{Latex[length=1.6mm]}, semithick, gray!70!black},
    rateA/.style={font=\scriptsize\bfseries, red!55!black},
    rateB/.style={font=\scriptsize\bfseries, blue!55!black},
    panelttl/.style={font=\small\bfseries},
    panelsub/.style={font=\scriptsize\itshape, gray!45!black},
    plaintiffbox/.style={draw, dashed, rounded corners=3pt, inner sep=4pt, gray!55!black},
    flatbar/.style={fill=gray!30, draw=gray!55!black},
    survivebar/.style={fill=red!22, draw=red!55!black},
]

%==================================================================
% PANEL A -- Unadjusted procedural ladder
%==================================================================
\begin{scope}[xshift=0cm]
  \node[panelttl, anchor=north west] at (-0.7,4.30) {A. Unadjusted ladder};
  \node[panelsub, anchor=north west] at (-0.7,3.95) {specialist (red) vs. other (blue)};

  \node[spec]    (a_def_s) at (0.55,2.75) {Default};
  \node[nonspec] (a_def_n) at (2.45,2.75) {Default};

  \node[spec]    (a_wri_s) at (0.55,1.85) {Writ};
  \node[nonspec] (a_wri_n) at (2.45,1.85) {Writ};

  \node[spec]    (a_ser_s) at (0.55,0.95) {Served writ};
  \node[nonspec] (a_ser_n) at (2.45,0.95) {Served writ};

  \node[rateA, anchor=east] at ([xshift=-2pt]a_def_s.west) {52.4\%};
  \node[rateB, anchor=west] at ([xshift=2pt]a_def_n.east) {52.8\%};
  \node[rateA, anchor=east] at ([xshift=-2pt]a_wri_s.west) {51.0\%};
  \node[rateB, anchor=west] at ([xshift=2pt]a_wri_n.east) {28.4\%};
  \node[rateA, anchor=east] at ([xshift=-2pt]a_ser_s.west) {21.0\%};
  \node[rateB, anchor=west] at ([xshift=2pt]a_ser_n.east) {8.6\%};

  \draw[arr] (a_def_s) -- (a_wri_s);
  \draw[arr] (a_wri_s) -- (a_ser_s);
  \draw[arr] (a_def_n) -- (a_wri_n);
  \draw[arr] (a_wri_n) -- (a_ser_n);

  \node[font=\scriptsize, align=center] at (1.50,0.10) {gap opens at writ \& served writ};
\end{scope}

%==================================================================
% PANEL B -- Within-plaintiff and within-plaintiff-property
%==================================================================
\begin{scope}[xshift=6.2cm]
  \node[panelttl, anchor=north west] at (-0.3,4.30) {B. Within-unit comparison};
  \node[panelsub, anchor=north west] at (-0.3,3.95) {same plaintiff and property};

  \node[plaintiffbox] (bbox) at (1.75,2.05) [minimum width=4.0cm, minimum height=3.0cm] {};

  \begin{scope}[xshift=0.2cm, yshift=1.05cm]
    \draw[->, semithick, gray!55!black] (0,0) -- (3.35,0);
    \draw[->, semithick, gray!55!black] (0,0) -- (0,2.20);
    \node[font=\tiny, gray!55!black, rotate=90, anchor=south] at (-0.22,1.05) {coef.\,(pp)};
    \draw[gray!40, dashed, thin] (0,1.05) -- (3.30,1.05);
    \node[font=\tiny, gray!55!black, anchor=east] at (-0.04,1.05) {0};

    \fill[flatbar] (0.20,1.05) rectangle (0.65,1.16);
    \node[font=\tiny, anchor=south] at (0.425,1.16) {+0.9};
    \node[font=\tiny, anchor=north, gray!55!black] at (0.425,-0.04) {def};

    \fill[flatbar] (0.85,1.05) rectangle (1.30,1.01);
    \node[font=\tiny, anchor=north, gray!55!black] at (1.075,-0.04) {judg};

    \fill[survivebar] (1.50,1.05) rectangle (1.95,0.63);
    \node[font=\tiny, anchor=north, red!55!black] at (1.725,0.63) {$-3.2$};
    \node[font=\tiny, anchor=north, gray!55!black] at (1.725,-0.04) {writ};

    \fill[flatbar] (2.15,1.05) rectangle (2.60,0.89);
    \node[font=\tiny, anchor=north, gray!55!black] at (2.375,-0.04) {served};

    \fill[flatbar] (2.80,1.05) rectangle (3.25,1.02);
    \node[font=\tiny, anchor=north, gray!55!black] at (3.025,-0.04) {fee};
  \end{scope}

  \node[font=\tiny\itshape, anchor=north east, gray!55!black] at (3.65,3.45) {plaintiff $\times$ property fixed};

  \node[font=\scriptsize, align=center] at (1.75,0.10) {no broad adverse premium};
\end{scope}

%==================================================================
% PANEL C -- Scale and sequence
%==================================================================
\begin{scope}[xshift=11.5cm]
  \node[panelttl, anchor=north west] at (-0.3,4.30) {C. Scale and sequence};
  \node[panelsub, anchor=north west] at (-0.3,3.95) {after adoption or top-10 switch};

  \draw[->, semithick, gray!55!black] (0.30,0.95) -- (3.50,0.95);
  \node[font=\tiny, gray!55!black, anchor=north east] at (3.50,0.93) {time};
  \draw[->, semithick, gray!55!black] (0.30,0.95) -- (0.30,3.30);

  \draw[dashed, semithick, gray!55!black] (1.85,0.95) -- (1.85,3.05);
  \node[font=\tiny\bfseries, anchor=south, gray!35!black] at (1.85,3.05) {adopt / switch};

  \draw[very thick, green!45!black] (0.40,1.65) -- (1.85,1.65);
  \draw[very thick, green!45!black] (1.85,1.65) -- (3.40,2.60);
  \node[font=\tiny, anchor=south, green!45!black] at (1.10,1.70) {filings};

  \draw[very thick, green!30!black, densely dashed] (0.40,1.30) -- (1.85,1.30);
  \draw[very thick, green!30!black, densely dashed] (1.85,1.30) -- (3.40,2.15);
  \node[font=\tiny, anchor=north, green!30!black] at (1.10,1.30) {buildings};

  \draw[<->, thin, green!45!black] (3.48,1.65) -- (3.48,2.60);
  \node[font=\scriptsize\bfseries, anchor=west, green!35!black] at (3.50,2.15) {$+2$--$5\%$};

  \node[font=\scriptsize, align=center] at (1.85,0.45) {pre-trends not flat (SI)};
\end{scope}

\end{tikzpicture}
\caption{Three components of the single-appearance result. (A) Unadjusted ladder: specialist (red) and non-specialist (blue) cases end in default at nearly the same rate ($\sim$52\%), but specialist cases are much more likely to advance to writ-of-possession (51.0\% vs.\ 28.4\%) and served writ (21.0\% vs.\ 8.6\%). (B) Within-unit comparison: once the same plaintiff is followed at the same property, the adverse-outcome premium disappears; default, judgment, served writ, and fee share are statistically null, and writ issuance is in fact \emph{lower} by 3.2 percentage points. (C) Scale and sequence: after a plaintiff adopts or switches into high-volume counsel, monthly filings and the number of distinct buildings reached both rise by roughly 2--5\%; pre-treatment trends are not flat, so the causal interpretation is qualified (see Supplementary Information).}
\label{fig:three_panel_schema}
\end{figure}

Figure~\ref{fig:three_panel_schema} summarizes the three components of the result that follow: unadjusted enforcement asymmetries, the within-unit comparisons that absorb them, and the scale-and-sequence pattern that survives.

\subsection{Specialist cases are enforcement-heavy in the unadjusted data}

In the residential single-appearance sample, specialist plaintiff-side counsel handles 212{,}074 cases, while other, non-specialist, or missing plaintiff-side counsel handles 184{,}089 cases. Unadjusted specialist cases are not more default-heavy: default is 52.4\% under specialist counsel and 52.8\% otherwise. The unadjusted differences instead appear at later procedural stages. Writ issuance is 51.0\% under specialist counsel and 28.4\% otherwise. Served writ is 21.0\% under specialist counsel and 8.6\% otherwise. JBA is also higher under specialist counsel, while judgment for plaintiff is lower.

The unadjusted ladder therefore rules out a simple descriptive claim that specialist attorneys raise every adverse endpoint. Specialist cases are more enforcement-heavy on the writ and served-writ margins, but not more default-heavy. The unadjusted patterns motivate the within-unit tests that follow.

\begin{figure}[H]
    \centering
    \includegraphics[width=\linewidth]{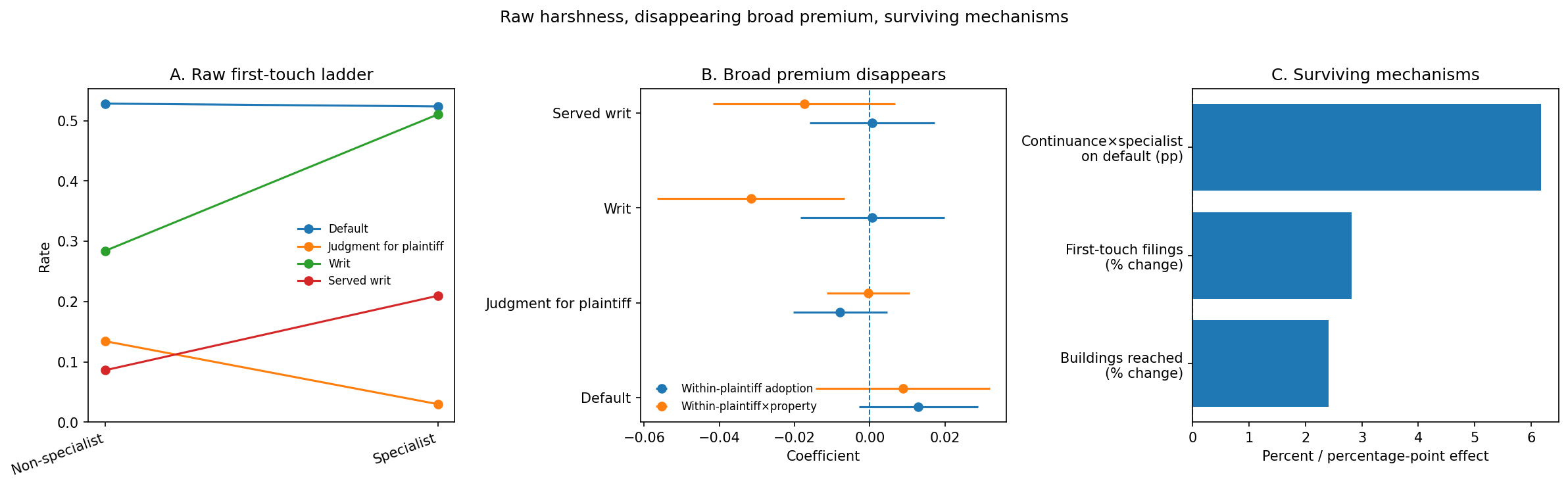}
    \caption{Synthesis of the single-appearance tenant results. Panel A shows the unadjusted outcome ladder. Specialist cases are much more writ- and served-writ-heavy, but not more default-heavy. Panel B reports selected within-unit estimates, where a broad adverse-endpoint premium is not supported. Panel C reports post-treatment scale estimates for top-decile adoption and top-10 switching.}
    \label{fig:killer_synthesis}
\end{figure}

\subsection{Within-unit estimates do not support a broad specialist endpoint premium}

The within-plaintiff results do not show a broad adverse specialist premium. Specialist use is associated with a small positive but statistically imprecise default coefficient (+0.0129, $p=0.111$), near-zero writ (+0.0007, $p=0.945$), and near-zero served writ (+0.0006, $p=0.941$). Judgment for plaintiff is negative but imprecise. Fee share is lower by 0.0052 ($p=0.0025$), and award-over-sought is positive but imprecise.

The within-plaintiff--property design pushes further against the unadjusted enforcement story. Holding the same plaintiff and property fixed, specialist use is not associated with higher default (+0.0088, $p=0.458$), judgment for plaintiff ($-0.0005$, $p=0.930$), served writ ($-0.0174$, $p=0.158$), fee share, or award-over-sought. Writ issuance is lower by 3.16 percentage points ($p=0.0128$). The strongest within-unit comparison therefore does not support the view that specialist counsel mechanically raises the main single-appearance endpoints once plaintiff--property structure is held fixed.

\begin{table}[t]
\centering
\caption{Within-unit specialist estimates in residential single-appearance cases.}
\label{tab:within_unit}
\small
\begin{tabular}{p{0.26\linewidth}p{0.23\linewidth}p{0.23\linewidth}p{0.22\linewidth}}
\toprule
Outcome & Within-plaintiff estimate & Within-plaintiff--property estimate & Interpretation \\
\midrule
Default & $+0.0129$ ($p=0.111$) & $+0.0088$ ($p=0.458$) & no stable default premium \\
Judgment for plaintiff & $-0.0079$ ($p=0.218$) & $-0.0005$ ($p=0.930$) & no stable judgment premium \\
Writ & $+0.0007$ ($p=0.945$) & $-0.0316$ ($p=0.0128$) & no increase; lower within property \\
Served writ & $+0.0006$ ($p=0.941$) & $-0.0174$ ($p=0.158$) & no stable served-writ premium \\
Fee share & $-0.0052$ ($p=0.0025$) & $-0.0017$ ($p=0.419$) & fee share does not rise \\
Award / amount sought & $+0.0517$ ($p=0.395$) & $+0.0030$ ($p=0.967$) & no stable award inflation \\
\bottomrule
\end{tabular}
\end{table}

Across the stable-attorney window and later windows, alternative high-volume definitions do not recover a universal adverse endpoint premium. The repeated signal is selective: JBA and fee patterns change in some within-unit specifications, but default and served-writ premiums are not stable.

\subsection{Adoption and switching increase filing scale}

The clearest positive result is on plaintiff-month scale. In the top-decile adoption design, adoption is followed by 2.33\% more monthly residential lifetime single-appearance filings ($p<0.001$) and 2.75\% more buildings reached ($p<0.001$) during months 1--12 after adoption. These estimates imply about 0.55 additional single-appearance filings and 0.60 additional buildings reached per adopting plaintiff-year.

The direct top-10 switch design is larger. Among plaintiffs with sufficient pre-switch histories, moving from non-top-10 to top-10 counsel is followed by 5.00\% more monthly single-appearance filings ($p<0.001$) and 4.86\% more buildings reached ($p<0.001$). Package-level validation at quarter resolution points in the same direction: the comparable post-1--4-quarter event-time means are +1.40\% for top-decile filings, +1.25\% for top-decile buildings, +1.82\% for top-10 switch filings, and +1.74\% for top-10 switch buildings. These validation estimates are used as sign-and-magnitude checks, not as formal package inference.

Pre-treatment trends are not flat, so the unadjusted post-treatment estimates should not be read as randomized treatment effects. Pretrend-adjusted estimates appear in the Supplementary Information and replication files as sensitivity quantities. The statistically significant pattern is a post-treatment scale increase; its causal interpretation is qualified by pretrend behavior.

\begin{table}[t]
\centering
\caption{Scale estimates in residential single-appearance cases.}
\label{tab:scale_results}
\small
\begin{tabular}{p{0.36\linewidth}p{0.18\linewidth}p{0.18\linewidth}p{0.21\linewidth}}
\toprule
Design and outcome & Monthly post-treatment effect & Package validation & Interpretation \\
\midrule
Top-decile adoption: filings & $+2.33\%$ ($p<0.001$) & $+1.40\%$ & larger monthly filing volume \\
Top-decile adoption: buildings & $+2.75\%$ ($p<0.001$) & $+1.25\%$ & broader building reach \\
Top-10 switch: filings & $+5.00\%$ ($p<0.001$) & $+1.82\%$ & larger post-switch filing volume \\
Top-10 switch: buildings & $+4.86\%$ ($p<0.001$) & $+1.74\%$ & broader post-switch building reach \\
Top-decile filings per adopting plaintiff-year & $+0.55$ & -- & translated scale \\
Top-decile buildings per adopting plaintiff-year & $+0.60$ & -- & translated scale \\
\bottomrule
\end{tabular}
\vspace{0.3em}
\raggedright\footnotesize Notes: Monthly effects average months 1--12 after adoption or switch. Package validation reports the post-1--4-quarter event-time mean and is used only as a sign-and-magnitude check. Pretrend-adjusted estimates are positive but not statistically sharp and are reported in the Supplementary Information.
\end{table}

\subsection{Rank-threshold, sequence, and robustness results identify selective mechanisms}

The rank-10 threshold design provides local evidence that the top-10 category is not merely a label. Cases handled by attorneys just inside the prior-year top-10 threshold are more default-heavy than cases just outside it. The main rank-clustered RD estimate is +3.70 percentage points for default ($p=0.006$). Fee share moves in the opposite direction, with a main RD coefficient of $-3.22$ percentage points ($p=0.014$). Donut RD estimates that drop ranks 10 and 11 are stronger: +6.08 percentage points for default ($p<0.001$), +7.97 percentage points for writ ($p<0.001$), +6.70 percentage points for served writ ($p<0.001$), and $-4.77$ percentage points for fee share ($p=0.004$). Because the running variable is an integer rank and rank is mechanically tied to prior-year volume, these estimates are local threshold evidence rather than a global treatment effect.

Sequence and robustness results reinforce the same interpretation. In the default-continuance model, the specialist main effect is negative ($-0.129$, $p=0.0029$), and the specialist$\times$continuance interaction is positive (+0.0618, $p<0.001$). Continuances under specialist counsel are therefore more tightly linked to default, without implying a uniform default premium across all specialist cases. In the within-plaintiff fee model, fee share falls by 0.0052 ($p=0.0025$), while award-over-sought is statistically null. In the follow-up-balanced sample through 2018, within-plaintiff writ issuance is positive (+0.0205, $p=0.041$), but fee share remains negative ($-0.0055$, $p=0.003$). In a plaintiff--property DDD, the only significant differential exposure effect is on judgment by agreement (+4.24 percentage points, $p=0.030$). Negative-control random-adoption placebos are centered near zero, supporting the view that the scaling estimates are not a mechanical artifact of the panel construction.

The reverse-direction attorney-exit design is reported as a diagnostic. The exit algorithm detects 14 top-10 attorney exits, of which 11 contribute support to the event-level bootstrap path. Plaintiffs losing dominant top-10 counsel show a negative post-exit point estimate, but the post-1--12 aggregate is not statistically significant. The result points in the expected direction, but it is not used as a headline statistical finding.

\begin{table}[t]
\centering
\caption{Significant triangulation, robustness, and sequence results.}
\label{tab:triangulation}
\small
\begin{tabular}{p{0.39\linewidth}p{0.20\linewidth}p{0.14\linewidth}p{0.19\linewidth}}
\toprule
Finding & Estimate & Inference & Interpretation \\
\midrule
Rank-10 RD, default & $+3.70$ pp & $p=0.006$ & local top-10 threshold difference \\
Rank-10 RD, fee share & $-3.22$ pp & $p=0.014$ & no local fee premium \\
Donut RD, default / writ / served writ & $+6.08$ / $+7.97$ / $+6.70$ pp & all $p<0.001$ & stronger local enforcement differences \\
Donut RD, fee share & $-4.77$ pp & $p=0.004$ & no fee-share increase at the threshold \\
Specialist$\times$continuance & $+0.0618$ & $p<0.001$ & continuances under specialists are more default-linked \\
Within-plaintiff fee share & $-0.0052$ & $p=0.0025$ & fee share falls within plaintiff \\
Follow-up-balanced writ, within plaintiff & $+0.0205$ & $p=0.041$ & longer follow-up recovers a writ margin \\
Follow-up-balanced fee share, within plaintiff & $-0.0055$ & $p=0.003$ & no fee premium under longer follow-up \\
Plaintiff--property DDD, JBA & $+0.0424$ & $p=0.030$ & specialist-exposed properties show more agreement activity \\
\bottomrule
\end{tabular}
\end{table}

\paragraph{Synthesis.}
The significant findings cohere around scale and sequence. Specialist counsel is associated with higher monthly filing volume, broader building reach, local differences in default and enforcement at the top-10 threshold, and a stronger continuance--default link. Within-unit comparisons do not yield a universal adverse-outcome premium, and the causal claim is restrained by non-flat pre-treatment trends and imprecise attorney-exit evidence.

\section{Discussion, conclusion, and limitations}\label{sec:discussion}

\subsection{Principal contribution}

High-volume plaintiff-side counsel shapes the largest defensible population of single-appearance tenants in the Philadelphia eviction docket. The evidence does not show that specialist attorneys worsen every case outcome; it shows that high-volume counsel is tied to filing scale, building reach, and procedural sequence.

The descriptive and within-unit results narrow the mechanism. Unadjusted specialist cases are enforcement-heavy, but within-plaintiff and within-plaintiff--property models do not show a stable premium on every adverse case outcome. The staggered adoption and switch estimates identify filing-scale changes. The rank-threshold and sequence results identify local enforcement and procedural-routing differences. Together, these findings describe a legal mechanism that organizes court entry rather than a case-level treatment that uniformly worsens outcomes.

\subsection{Statistical implications}

The main statistical lesson is that high-volume legal intermediation should be studied as a dynamic panel problem. Plaintiffs adopt or switch counsel at different times, attorney categories depend on prior-year volume, and outcomes unfold across legal stages. A single two-way fixed-effect adoption coefficient is not sufficient for this structure. The strategy adopted here combines staggered adoption comparisons, within-unit fixed effects, rank-threshold comparisons, placebo checks, and follow-up-balanced robustness.

Significant and null findings must be read together. Significant post-treatment scale effects show that filing volume and building reach rise after adoption or switching. Significant threshold and sequence results show selective procedural differences. At the same time, non-flat pre-treatment trends and imprecise attorney-exit estimates prevent a stronger causal claim. The resulting characterization of plaintiff-side legal specialization in one-shot court entry is statistical rather than mechanistic, and disciplined rather than universal.

\subsection{Epistemic scope and limitations}

The claims remain narrower than a fully causal treatment effect. First, single-appearance status is an administrative docket classification. It does not identify all housing conflict, all informal displacement, or all removals. It also is not an individual treatment response. The adoption and switch estimands concern plaintiff-month initiation of residential lifetime single-appearance filings and buildings reached.

Second, specialist adoption and top-10 switching are not random. The staggered designs compare adopting or switching plaintiffs with not-yet-treated and never-treated plaintiffs, but pre-treatment trends reject in several specifications. For that reason, pretrend-adjusted estimates are reported as sensitivity checks and are not used as significant positive findings. Third, the attorney-exit design points in the expected reverse direction but is too imprecise to stand alone. Fourth, fee, judge, and within-property modules operate on smaller identified samples because they require modern fields or within-unit variation.

These limitations define the scope of the result. The evidence supports an organizational account of high-volume plaintiff-side counsel: specialists expand and route single-appearance court entry, but the data do not support a claim that counsel uniformly worsens every endpoint or causes particular tenants to become single-appearance.

\subsection{Conclusion}

Single-appearance eviction cases are organized through repeat plaintiffs, specialized counsel, and procedural routines. In Philadelphia, high-volume plaintiff-side counsel is most visible as a mechanism of filing scale and procedural sequence: it expands the geographic and quantitative reach of filing activity and shifts the procedural path through the docket, while within-unit comparisons reject a uniform adverse-outcome premium. The implication is both statistical and institutional: legal intermediation shapes how single-appearance court entry is produced, and the strongest evidence concerns plaintiff filing behavior rather than tenant-level case outcomes.

\section*{Acknowledgments}

The dataset is available at \url{https://docs.philalegal.org/index.php/s/w9IQZrb8eDqXJkU}. We thank Jonathan Pyle at Philadelphia Legal Assistance for providing the Philadelphia Municipal Court records.

\bibliographystyle{apalike}
\bibliography{Eviction}

@article{hartman2003evictions,
  title={Evictions: The hidden housing problem},
  author={Hartman, Chester and Robinson, David},
  journal={Housing Policy Debate},
  volume={14},
  number={4},
  pages={461--501},
  year={2003},
  publisher={Taylor \& Francis}
}

@article{de2020two,
  title={Two-way fixed effects estimators with heterogeneous treatment effects},
  author={De Chaisemartin, Cl{\'e}ment and d’Haultfoeuille, Xavier},
  journal={American economic review},
  volume={110},
  number={9},
  pages={2964--2996},
  year={2020},
  publisher={American Economic Association 2014 Broadway, Suite 305, Nashville, TN 37203}
}

@article{sun2021estimating,
  title={Estimating dynamic treatment effects in event studies with heterogeneous treatment effects},
  author={Sun, Liyang and Abraham, Sarah},
  journal={Journal of econometrics},
  volume={225},
  number={2},
  pages={175--199},
  year={2021},
  publisher={Elsevier}
}

@article{goodman2021difference,
  title={Difference-in-differences with variation in treatment timing},
  author={Goodman-Bacon, Andrew},
  journal={Journal of econometrics},
  volume={225},
  number={2},
  pages={254--277},
  year={2021},
  publisher={Elsevier}
}

@article{callaway2021difference,
  title={Difference-in-differences with multiple time periods},
  author={Callaway, Brantly and Sant’Anna, Pedro HC},
  journal={Journal of econometrics},
  volume={225},
  number={2},
  pages={200--230},
  year={2021},
  publisher={Elsevier}
}

@article{desmond2012eviction,
  title={Eviction and the reproduction of urban poverty},
  author={Desmond, Matthew},
  journal={American journal of sociology},
  volume={118},
  number={1},
  pages={88--133},
  year={2012},
  publisher={University of Chicago Press Chicago, IL}
}

@article{desmond2015eviction,
  title={Eviction's fallout: housing, hardship, and health},
  author={Desmond, Matthew and Kimbro, Rachel Tolbert},
  journal={Social forces},
  volume={94},
  number={1},
  pages={295--324},
  year={2015},
  publisher={Oxford University Press}
}

@article{gromis2022estimating,
  title={Estimating eviction prevalence across the United States},
  author={Gromis, Ashley and Fellows, Ian and Hendrickson, James R and Edmonds, Lavar and Leung, Lillian and Porton, Adam and Desmond, Matthew},
  journal={Proceedings of the National Academy of Sciences},
  volume={119},
  number={21},
  pages={e2116169119},
  year={2022},
  publisher={National Academy of Sciences}
}

@article{graetz2023comprehensive,
  title={A comprehensive demographic profile of the US evicted population},
  author={Graetz, Nick and Gershenson, Carl and Hepburn, Peter and Porter, Sonya R and Sandler, Danielle H and Desmond, Matthew},
  journal={Proceedings of the National Academy of Sciences},
  volume={120},
  number={41},
  pages={e2305860120},
  year={2023},
  publisher={National Academy of Sciences}
}

@article{collinson2024eviction,
  title={Eviction and poverty in American cities},
  author={Collinson, Robert and Humphries, John Eric and Mader, Nicholas and Reed, Davin and Tannenbaum, Daniel and Van Dijk, Winnie},
  journal={The Quarterly Journal of Economics},
  volume={139},
  number={1},
  pages={57--120},
  year={2024},
  publisher={Oxford University Press}
}

@article{nelson2021evictions,
  title={Evictions: The comparative analysis problem},
  author={Nelson, Kyle and Garboden, Philip and McCabe, Brian J and Rosen, Eva},
  journal={Housing Policy Debate},
  volume={31},
  number={3-5},
  pages={696--716},
  year={2021},
  publisher={Taylor \& Francis}
}

@article{porton2021inaccuracies,
  title={Inaccuracies in eviction records: Implications for renters and researchers},
  author={Porton, Adam and Gromis, Ashley and Desmond, Matthew},
  journal={Housing Policy Debate},
  volume={31},
  number={3-5},
  pages={377--394},
  year={2021},
  publisher={Taylor \& Francis}
}

@article{summers2025pathways,
  title={Pathways to eviction},
  author={Summers, Nicole and Steil, Justin},
  journal={Law \& Social Inquiry},
  volume={50},
  number={1},
  pages={129--169},
  year={2025},
  publisher={Cambridge University Press}
}

@article{galanter1974haves,
  title={Why the “haves” come out ahead: Speculations on the limits of legal change},
  author={Galanter, Marc},
  journal={Law \& society review},
  volume={9},
  number={1},
  pages={95--160},
  year={1974},
  publisher={Cambridge University Press \& Assessment}
}

@article{sudeall2021praxis,
  title={Praxis and paradox: Inside the Black Box of eviction court},
  author={Sudeall, Lauren and Pasciuti, Daniel},
  journal={Vand. L. Rev.},
  volume={74},
  pages={1365},
  year={2021},
  publisher={HeinOnline}
}

@article{engler2010connecting,
  title={Connecting self-representation to civil Gideon: What existing data reveal about when counsel is most needed},
  author={Engler, Russell},
  journal={Fordham Urb. LJ},
  volume={37},
  pages={37},
  year={2010},
  publisher={HeinOnline}
}

@article{sabbeth2022eviction,
  title={Eviction courts},
  author={Sabbeth, Kathryn A},
  journal={U. St. Thomas LJ},
  volume={18},
  pages={359},
  year={2022},
  publisher={HeinOnline}
}

@article{aizman2025shadow,
  title={Shadow players of the eviction crisis: identifying and characterizing professional evicting attorneys in Massachusetts},
  author={Aizman, Asya and Huntley, Eric Robsky},
  journal={Housing Studies},
  pages={1--24},
  year={2025},
  publisher={Taylor \& Francis}
}

@article{summers2024evicted,
  title={Evicted by Default},
  author={Summers, Nicole and Steil, Justin},
  journal={Conn. L. Rev.},
  volume={57},
  pages={1233},
  year={2024},
  publisher={HeinOnline}
}

@article{cassidy2023effects,
  title={The effects of legal representation on tenant outcomes in housing court: Evidence from New York City’s Universal Access program},
  author={Cassidy, Mike and Currie, Janet},
  journal={Journal of Public Economics},
  volume={222},
  pages={104844},
  year={2023},
  publisher={Elsevier}
}

@article{garboden2019serial,
  title={Serial filing: How landlords use the threat of eviction},
  author={Garboden, Philip ME and Rosen, Eva},
  journal={City \& Community},
  volume={18},
  number={2},
  pages={638--661},
  year={2019},
  publisher={SAGE Publications Sage CA: Los Angeles, CA}
}

@article{leung2021serial,
  title={Serial eviction filing: Civil courts, property management, and the threat of displacement},
  author={Leung, Lillian and Hepburn, Peter and Desmond, Matthew},
  journal={Social Forces},
  volume={100},
  number={1},
  pages={316--344},
  year={2021},
  publisher={Oxford University Press}
}

@article{immergluck2020evictions,
  title={Evictions, large owners, and serial filings: Findings from Atlanta},
  author={Immergluck, Dan and Ernsthausen, Jeff and Earl, Stephanie and Powell, Allison},
  journal={Housing Studies},
  volume={35},
  number={5},
  pages={903--924},
  year={2020},
  publisher={Taylor \& Francis}
}

@article{hepburn2020racial,
  title={Racial and gender disparities among evicted Americans},
  author={Hepburn, Peter and Louis, Renee and Desmond, Matthew},
  journal={Sociological Science},
  volume={7},
  pages={649--662},
  year={2020}
}

@article{gomory2023racially,
  title={The racially disparate influence of filing fees on eviction rates},
  author={Gomory, Henry and Massey, Douglas S and Hendrickson, James R and Desmond, Matthew},
  journal={Housing Policy Debate},
  volume={33},
  number={6},
  pages={1463--1483},
  year={2023},
  publisher={Taylor \& Francis}
}

@article{ajayi2026landlord,
  title={Landlord responsiveness to eviction filing fees: evidence from northern New England},
  author={Ajayi, Oluwafisayo and Hobbs, Kelsi G and Wibabara, Eliane},
  journal={Regional Studies, Regional Science},
  volume={13},
  number={1},
  pages={2620937},
  year={2026},
  publisher={Taylor \& Francis}
}

@article{brito2022racial,
  title={Racial capitalism in the civil courts},
  author={Brito, Tonya L and Sabbeth, Kathryn A and Steinberg, Jessica K and Sudeall, Lauren},
  journal={Colum. L. Rev.},
  volume={122},
  pages={1243},
  year={2022},
  publisher={HeinOnline}
}

@article{summers2023civil,
  title={Civil probation},
  author={Summers, Nicole},
  journal={Stan. L. Rev.},
  volume={75},
  pages={847},
  year={2023},
  publisher={HeinOnline}
}

@article{summers2026settlements,
  title={Settlements of Adhesion},
  author={Summers, Nicole},
  journal={University of Chicago Law Review},
  volume={93},
  number={1},
  year={2026}
}

@article{kleysteuber2006tenant,
  title={Tenant screening thirty years later: A statutory proposal to protect public records},
  author={Kleysteuber, Rudy},
  journal={Yale LJ},
  volume={116},
  pages={1344},
  year={2006},
  publisher={HeinOnline}
}

@article{eisenberg2024record,
  title={Record Costs: Collateral Consequences of Eviction Court Filings in Pennsylvania},
  author={Eisenberg, Alexa and Brantley, Kate},
  journal={Ann Arbor: University of Michigan, Housing Solutions for Health Equity},
  year={2024}
}

@article{brantley2025record,
  title={Record costs: examining the impact of eviction filings for tenants and their families},
  author={Brantley, Kate and Eisenberg, Alexa and Mehdipanah, Roshanak},
  journal={Housing Studies},
  pages={1--27},
  year={2025},
  publisher={Taylor \& Francis}
}

@article{hepburn2023protecting,
  title={Protecting the most vulnerable: policy response and eviction filing patterns during the COVID-19 pandemic},
  author={Hepburn, Peter and Haas, Jacob and Graetz, Nick and Louis, Renee and Rutan, Devin Q and Alexander, Anne Kat and Rangel, Jasmine and Jin, Olivia and Benfer, Emily and Desmond, Matthew},
  journal={RSF: The Russell Sage Foundation Journal of the Social Sciences},
  volume={9},
  number={3},
  pages={186--207},
  year={2023},
  publisher={RSF: The Russell Sage Foundation Journal of the Social Sciences}
}

@article{benfer2023covid,
  title={COVID-19 housing policy: State and federal eviction moratoria and supportive measures in the United States during the pandemic},
  author={Benfer, Emily A and Koehler, Robert and Mark, Alyx and Nazzaro, Valerie and Alexander, Anne Kat and Hepburn, Peter and Keene, Danya E and Desmond, Matthew},
  journal={Housing Policy Debate},
  volume={33},
  number={6},
  pages={1390--1414},
  year={2023},
  publisher={Taylor \& Francis}
}

@article{zainulbhai2022informal,
  title={Informal evictions: Measuring displacement outside the courtroom},
  author={Zainulbhai, Sabiha and Daly, Nora},
  year={2022},
  publisher={< bound method Organization. get\_name\_with\_acronym of< Organization: New~…}
}

@article{shanahan2022judges,
  title={Judges in Lawyerless Courts},
  author={Shanahan, Colleen F and Carpenter, Anna E and Steinberg, Jessica and Mark, Alyx},
  journal={Georgetown Law Journal},
  pages={509},
  year={2022}
}

@book{desmond2017evicted,
  title={Evicted: Poverty and profit in the American city},
  author={Desmond, Matthew},
  year={2017},
  publisher={Crown}
}

@article{desmond2017gets,
  title={Who gets evicted? Assessing individual, neighborhood, and network factors},
  author={Desmond, Matthew and Gershenson, Carl},
  journal={Social science research},
  volume={62},
  pages={362--377},
  year={2017},
  publisher={Elsevier}
}

@article{greenberg2016discrimination,
  title={Discrimination in evictions: empirical evidence and legal challenges},
  author={Greenberg, Deena and Gershenson, Carl and Desmond, Matthew},
  journal={Harv. CR-CLL Rev.},
  volume={51},
  pages={115},
  year={2016},
  publisher={HeinOnline}
}

@article{desmond2015forced,
  title={Forced relocation and residential instability among urban renters},
  author={Desmond, Matthew and Gershenson, Carl and Kiviat, Barbara},
  journal={Social Service Review},
  volume={89},
  number={2},
  pages={227--262},
  year={2015},
  publisher={University of Chicago Press Chicago, IL}
}

@incollection{felstiner2017emergence,
  title={The emergence and transformation of disputes: Naming, blaming, claiming…},
  author={Felstiner, William LF and Abel, Richard L and Sarat, Austin},
  booktitle={Theoretical and Empirical Studies of Rights},
  pages={255--306},
  year={2017},
  publisher={Routledge}
}

@article{mnookin1978bargaining,
  title={Bargaining in the shadow of the law: The case of divorce},
  author={Mnookin, Robert H and Kornhauser, Lewis},
  journal={Yale lJ},
  volume={88},
  pages={950},
  year={1978},
  publisher={HeinOnline}
}

@article{priest1984selection,
  title={The selection of disputes for litigation},
  author={Priest, George L and Klein, Benjamin},
  journal={The journal of legal studies},
  volume={13},
  number={1},
  pages={1--55},
  year={1984},
  publisher={The University of Chicago Law School}
}

@article{sandefur2015elements,
  title={Elements of professional expertise: Understanding relational and substantive expertise through lawyers’ impact},
  author={Sandefur, Rebecca L},
  journal={American Sociological Review},
  volume={80},
  number={5},
  pages={909--933},
  year={2015},
  publisher={Sage Publications Sage CA: Los Angeles, CA}
}

@article{pistor2019code,
  title={The code of capital: How the law creates wealth and inequality},
  author={Pistor, Katharina},
  year={2019},
  publisher={Princeton University Press}
}

@article{rabin1983revolution,
  title={Revolution in residential landlord-tenant law: causes and consequences},
  author={Rabin, Edward H},
  journal={Cornell L. Rev.},
  volume={69},
  pages={517},
  year={1983},
  publisher={HeinOnline}
}

@article{seron2001impact,
  title={The impact of legal counsel on outcomes for poor tenants in New York City's housing court: results of a randomized experiment},
  author={Seron, Carroll and Van Ryzin, Gregg and Frankel, Martin},
  journal={Law \& Society Review},
  volume={35},
  number={2},
  pages={419--434},
  year={2001},
  publisher={Cambridge University Press \& Assessment}
}

@article{bezdek1991silence,
  title={Silence in the court: Participation and subordination of poor tenants' voices in legal process},
  author={Bezdek, Barbara},
  journal={Hofstra L. Rev.},
  volume={20},
  pages={533},
  year={1991},
  publisher={HeinOnline}
}

@article{ellen2021lawyers,
  title={Do lawyers matter? Early evidence on eviction patterns after the rollout of universal access to counsel in New York City},
  author={Ellen, Ingrid Gould and O’Regan, Katherine and House, Sophia and Brenner, Ryan},
  journal={Housing Policy Debate},
  volume={31},
  number={3-5},
  pages={540--561},
  year={2021},
  publisher={Taylor \& Francis}
}

@article{summers2022eviction,
  title={Eviction court displacement rates},
  author={Summers, Nicole},
  journal={Nw. UL REv.},
  volume={117},
  pages={287},
  year={2022},
  publisher={HeinOnline}
}

@article{harris2010drawing,
  title={Drawing blood from stones: Legal debt and social inequality in the contemporary United States},
  author={Harris, Alexes and Evans, Heather and Beckett, Katherine},
  journal={American journal of sociology},
  volume={115},
  number={6},
  pages={1753--1799},
  year={2010},
  publisher={The University of Chicago Press}
}

@article{summers1limits,
  author  = {Summers, Nicole},
  title   = {The Limits of Good Law: A Study of Housing Court Outcomes},
  journal = {University of Chicago Law Review},
  year    = {2020},
  volume={1},
  pages={145}
}

@article{papamichalis2026legal,
  title={Legal Infrastructure Organizes Eviction: Evidence from Philadelphia},
  author={Papamichalis, Marios and Ruane, Regina},
  journal={arXiv preprint arXiv:2604.21212},
  year={2026}
}

@article{hardaway2026courthouse,
  title={Courthouse funneling: how organizational mechanisms teach tenants to bargain in eviction court},
  author={Hardaway, N and Clair, Matthew},
  journal={Law \& Society Review},
  pages={1--28},
  year={2026},
  publisher={Cambridge University Press}
}

@article{fleming2023navigating,
  title={Navigating an overburdened courtroom: How inconsistent rules, shadow procedures, and social capital disadvantage tenants in eviction court},
  author={Fleming-Klink, Isaiah and McCabe, Brian J and Rosen, Eva},
  journal={City \& Community},
  volume={22},
  number={3},
  pages={220--245},
  year={2023},
  publisher={SAGE Publications Sage CA: Los Angeles, CA}
}

@article{kepes2024system,
  title={“The System Is So Messed up”: Neutrality and Efficiency in an Eviction Courtroom},
  author={Kepes, Jacob Scott and Kempler, Alex M},
  journal={Socius},
  volume={10},
  pages={23780231241286928},
  year={2024},
  publisher={SAGE Publications Sage CA: Los Angeles, CA}
}

@article{hanley2024power,
  title={Power in the court: Legal argumentation and the hierarchy of credibility in eviction hearings},
  author={Hanley, Caroline and Howell, Kathryn and Teresa, Benjamin},
  journal={Socius},
  volume={10},
  pages={23780231241266510},
  year={2024},
  publisher={SAGE Publications Sage CA: Los Angeles, CA}
}

@article{ryan2024buying,
  title={Buying Time},
  author={Ryan Jr, Christopher J and Armstrong, Cassie Chambers},
  journal={Geo. J. on Poverty L. \& Pol'y},
  volume={32},
  pages={381},
  year={2024},
  publisher={HeinOnline}
}

@article{ryan2025tale,
  title={A Tale of Two Landlords},
  author={Ryan Jr, Christopher J},
  journal={Ky. LJ},
  volume={114},
  pages={1},
  year={2025},
  publisher={HeinOnline}
}

@article{metcalfe2016role,
  title={The role of courtroom workgroups in felony case dispositions: An analysis of workgroup familiarity and similarity},
  author={Metcalfe, Christi},
  journal={Law \& Society Review},
  volume={50},
  number={3},
  pages={637--673},
  year={2016},
  publisher={Cambridge University Press \& Assessment}
}

@article{benfer2025descriptive,
  title={A descriptive analysis of tenant right to counsel law and praxis 2017--2024},
  author={Benfer, Emily A and Hepburn, Peter and Nazarro, Valerie and Robinson, Leah and Michener, Jamila and Keene, Danya E},
  journal={Housing policy debate},
  volume={35},
  number={3},
  pages={470--495},
  year={2025},
  publisher={Taylor \& Francis}
}

@article{von2025no,
  title={No Right to Counsel: Evictions, Administrative Burden, and Access to Civil Justice},
  author={von Geldern, Will and Martin, Karin D},
  journal={Stan. L. \& Pol'y Rev.},
  volume={36},
  pages={313},
  year={2025},
  publisher={HeinOnline}
}

\clearpage
\appendix
\counterwithin{table}{section}
\counterwithin{figure}{section}
\renewcommand{\thesection}{\Alph{section}}
\renewcommand{\thesubsection}{\thesection.\arabic{subsection}}
\renewcommand{\thetable}{\thesection.\arabic{table}}
\renewcommand{\thefigure}{\thesection.\arabic{figure}}

\section*{Supplementary Information}
\addcontentsline{toc}{section}{Supplementary Information}

This supplement documents the denominator, the year distribution of single-appearance cases, and robustness across sample windows and specialist definitions. A single-appearance case is a case in which the tenant key appears exactly once in the full observed 1969--2022 docket.

\section{Sample construction and single-appearance denominator}

\subsection{Docket accounting}

The single-appearance construction proceeds from the full tenant-key audit. Among rows with usable tenant keys, 517{,}272 unique tenant identities appear; 401{,}527 occur exactly once and 115{,}745 occur more than once. The residential analytic sample contains 396{,}163 single-appearance rows.

\begin{table}[H]
\centering
\caption{Single-appearance denominator audit.}
\label{tab:si_denominator}
\small
\begin{tabular}{p{0.50\linewidth}p{0.18\linewidth}p{0.25\linewidth}}
\toprule
Metric & Count & Interpretation \\
\midrule
Unique tenant keys & 517{,}272 & observed tenant identities with usable keys \\
Tenant keys seen exactly once & 401{,}527 & lifetime single-appearance tenants \\
Tenant keys seen more than once & 115{,}745 & repeat tenants \\
single-appearance rows, all types & 401{,}527 & one docket row per single-appearance tenant \\
Residential single-appearance rows & 396{,}163 & main analytic sample \\
Rows missing usable tenant key & 104 & excluded from tenant-lifetime construction \\
\bottomrule
\end{tabular}
\end{table}

\subsection{Year distribution}

\begin{figure}[H]
    \centering
    \includegraphics[width=0.92\linewidth]{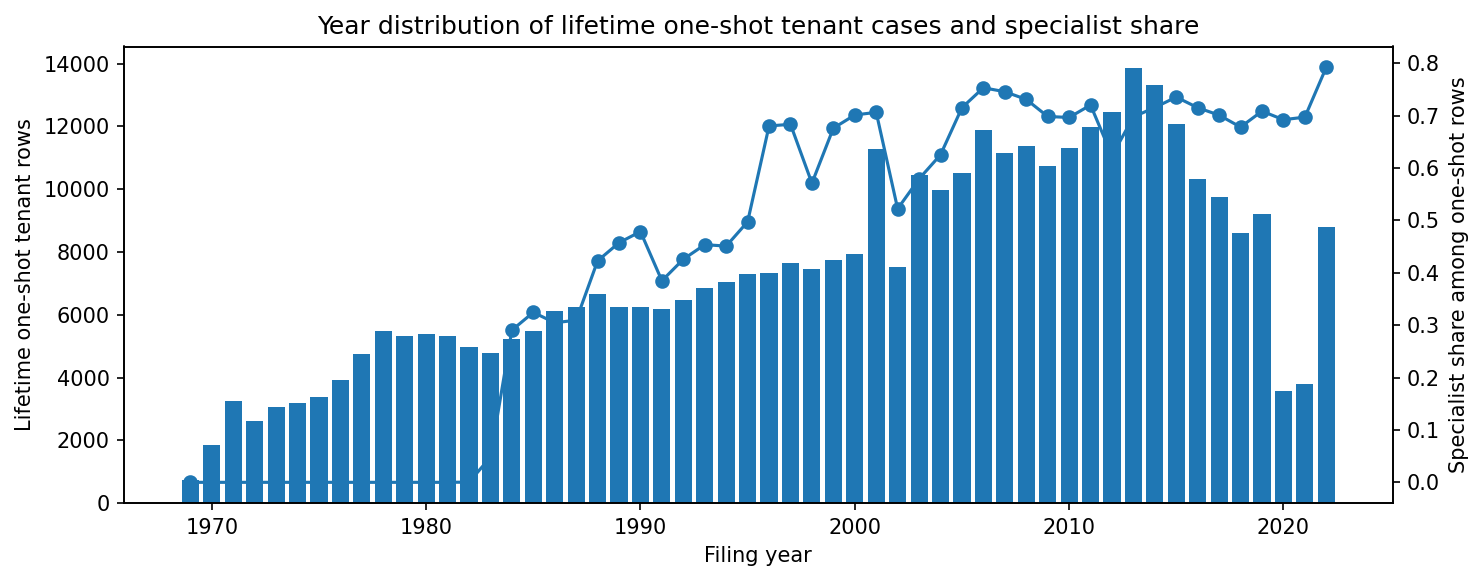}
    \caption{Year distribution of residential single-appearance tenant cases and specialist share. Bars show the number of residential single-appearance tenant rows by filing year. The line shows the specialist share among residential single-appearance rows.}
    \label{fig:si_year_distribution}
\end{figure}

Figure~\ref{fig:si_year_distribution} shows the year distribution of residential single-appearance tenant cases. The single-appearance tenant population is large throughout the modern docket. Specialist share rises sharply after stable plaintiff-side attorney identifiers become available and remains high in the modern period. 

\begin{table}[H]
\centering
\caption{Single-appearance sample counts by window.}
\label{tab:si_window_counts}
\small
\begin{tabular}{lrr}
\toprule
Sample window & Rows & Definition \\
\midrule
All years & 396{,}163 & residential lifetime single-appearance tenants \\
1983+ & 342{,}940 & stable plaintiff-side attorney window \\
1990+ & 302{,}195 & later cohort \\
2000+ & 231{,}872 & modern actor-rich window \\
2005+ & 184{,}696 & modern fee-rich window \\
\bottomrule
\end{tabular}
\end{table}

\section{Robustness across windows and specialist definitions}

\subsection{Top-decile specialist robustness}

The top-decile specialist definition produces the same qualitative result across windows. Specialist coefficients are not positive and stable for default or served writ. JBA is lower in the within-unit models, and fee share is lower or null. Table~\ref{tab:si_topdecile_robustness} reports selected estimates.

\begin{table}[H]
\centering
\caption{Selected top-decile robustness checks in single-appearance cases.}
\label{tab:si_topdecile_robustness}
\small
\begin{tabular}{p{0.26\linewidth}p{0.22\linewidth}p{0.20\linewidth}p{0.22\linewidth}}
\toprule
Specification & Outcome & Estimate & Interpretation \\
\midrule
1983+ stable-attorney window, within plaintiff & Default & $+0.0121$ ($p=0.136$) & no stable default premium \\
1983+ stable-attorney window, within plaintiff & JBA & $-0.0270$ ($p=0.007$) & lower agreement rate \\
1983+ stable-attorney window, within plaintiff & Served writ & $+0.0002$ ($p=0.984$) & no served-writ premium \\
1983+ stable-attorney window, within plaintiff & Fee share & $-0.0052$ ($p=0.002$) & lower fee share \\
1983+ stable-attorney window, within plaintiff--property & Default & $+0.0087$ ($p=0.458$) & no default premium \\
1983+ stable-attorney window, within plaintiff--property & JBA & $-0.0423$ ($p<0.001$) & lower agreement rate \\
1983+ stable-attorney window, within plaintiff--property & Served writ & $-0.0176$ ($p=0.152$) & no served-writ premium \\
1983+ stable-attorney window, within plaintiff--property & Fee share & $-0.0017$ ($p=0.407$) & null \\
\bottomrule
\end{tabular}
\end{table}

\subsection{Alternative high-volume definitions}

Alternative high-volume definitions do not recover a universal adverse endpoint premium. Under prior-year top-10 and prior-year top-5\% definitions, default remains null or negative, served writ remains null, and JBA remains lower in the within-unit models. The broad conclusion is therefore not an artifact of the top-decile cutoff.

\begin{table}[H]
\centering
\caption{Selected alternative specialist definitions, all-years single-appearance sample.}
\label{tab:si_alt_defs}
\small
\begin{tabular}{p{0.28\linewidth}p{0.24\linewidth}p{0.18\linewidth}p{0.22\linewidth}}
\toprule
Definition and design & Outcome & Estimate & Interpretation \\
\midrule
Prior-year top 10, within plaintiff & Default & $-0.0024$ ($p=0.696$) & no default premium \\
Prior-year top 10, within plaintiff & JBA & $-0.0191$ ($p=0.005$) & lower JBA \\
Prior-year top 10, within plaintiff--property & JBA & $-0.0273$ ($p=0.001$) & lower JBA within property \\
Prior-year top 5\%, within plaintiff & Default & $+0.0061$ ($p=0.377$) & no default premium \\
Prior-year top 5\%, within plaintiff & JBA & $-0.0392$ ($p<10^{-6}$) & lower JBA \\
Prior-year top 5\%, within plaintiff--property & JBA & $-0.0546$ ($p<10^{-7}$) & lower JBA within property \\
Prior-year top 5\%, within plaintiff & Fee share & $-0.0037$ ($p=0.006$) & lower fee share \\
\bottomrule
\end{tabular}
\end{table}

\section{Staggered adoption, ATTgt validation, exits, and rank-threshold robustness}

\subsection{Staggered adoption and package-validation outputs}

Table~\ref{tab:si_csdid_attgt} reports the scale estimates underlying the main text. The monthly CSDID estimates are the primary high-resolution estimates. The quarterly ATTgt validation agrees in sign over the post-1--4-quarter window. The package simple aggregation is not reported because it averages over a broader event grid and is not comparable to the near-post treatment window.

\begin{table}[h]
\centering
\caption{Monthly CSDID estimates and quarterly ATTgt validation.}
\label{tab:si_csdid_attgt}
\small
\begin{tabular}{p{0.31\linewidth}p{0.18\linewidth}p{0.22\linewidth}p{0.19\linewidth}}
\toprule
Design and outcome & Monthly unadjusted effect & Excess-over-pretrend & ATTgt post-1--4 quarter \\
\midrule
Top-decile adoption, filings & $+2.33\%$ & $+1.05\%$ ($p=0.721$) & $+1.40\%$ \\
Top-decile adoption, buildings & $+2.75\%$ & $+1.49\%$ ($p=0.548$) & $+1.25\%$ \\
Top-10 switch, filings & $+5.00\%$ & $+2.60\%$ ($p=0.304$) & $+1.82\%$ \\
Top-10 switch, buildings & $+4.86\%$ & $+2.48\%$ ($p=0.229$) & $+1.74\%$ \\
\bottomrule
\end{tabular}
\vspace{0.3em}
\raggedright\footnotesize Notes: Excess-over-pretrend estimates subtract the linear extrapolation of the lead path and are reported as sensitivity quantities when pre-treatment trends reject. ATTgt validation is reported as a sign-and-magnitude check, not as formal inference for the main monthly estimand.
\end{table}

\subsection{Attorney exits and rank-threshold diagnostics}

The exit algorithm detects 14 top-10 attorney exits, of which 11 contribute support to the event-level block-bootstrap path. The post-exit estimate is directionally negative ($-3.38\%$), but it is not statistically significant ($p=0.451$). The absorbed event-study pretrend diagnostic is $p=0.336$; the event-bootstrap heuristic diagnostic is $p=0.668$. The exit design is therefore reported as directional corroboration, not as a headline positive result.

The rank-threshold diagnostics are reported in the main text because they are statistically significant local comparisons around the prior-year top-10 cutoff. The density diagnostic shows modest imbalance at ranks 10 and 11, which is expected when rank is tied to prior-year volume; donut estimates dropping ranks 10 and 11 remain stronger than the main specification.

\subsection{Measurement, placebo, and follow-up diagnostics}

The tenant-key audit shows that the single-appearance denominator is not driven by obvious key-concentration artifacts. The top-100 tenant keys account for about 0.43\% of rows, and the maximum key frequency is 59. The hearing-day diagnostic shows that 99.085\% of residential single-appearance cases have a true hearing date and that only 0.94\% use the filing-date fallback.

Negative-control random-adoption placebos are centered near zero, with mean coefficient approximately 0.00009 and standard deviation approximately 0.00089 across 50 iterations. Follow-up-balanced models through 2018 show that within-plaintiff writ issuance becomes positive (+0.0205, $p=0.041$) while fee share remains negative ($-0.0055$, $p=0.003$). Within-plaintiff--property follow-up-balanced estimates remain mostly null.

\section{Synthesis}

The supplementary analyses support the scale-and-sequence interpretation. In the full single-appearance tenant population, unadjusted specialist cases are more writ- and served-writ-heavy, yet within-unit models reject a broad adverse-endpoint premium. Specialist adoption expands single-appearance filing and building reach, continuances under specialist counsel are more default-linked, and fee-share effects are negative or null rather than positive. The mechanism is organizational rather than uniformly punitive.

\end{document}